\documentstyle[aps,prb,multicol,epsf]{revtex}
\begin{document}
\title{Statistics of photodissociation spectra: nonuniversal properties}  
\author{Oded Agam} 
\address{The Racah Institute of Physics, The Hebrew University, 
Jerusalem 91904, Israel.} 
\maketitle 
\begin{abstract} 
We consider the two-point correlation function of the photodissociation 
cross section in molecules where the fragmentation process is indirect, passing
through resonances above the dissociation threshold. 
In the limit of overlapping resonances, a formula is derived, relating this 
correlation function to the behavior of the corresponding classical 
system. It is shown that nonuniversal features of the two-point correlation
function may have significant experimental manifestations.
\end{abstract} 

\begin{multicols}{2}

The photodissociation of molecules, such as the radicals $HO_2$ and $NO_2$, 
is an indirect process comprised of two steps. In the first step, 
a photon excites the molecule to an energy above the 
dissociation threshold. Then fragmentation proceeds via
energy redistribution within the vibrational degrees of freedom, or
tunneling from binding to unbinding energy surfaces of the adiabatic 
electronic potential\cite{book}. A barrier, which separates
quasi-stable states from continuum modes, hinders the immediate 
dissociation of the excited molecule. On these long lived resonances, 
the dynamics of the system is chaotic. Therefore, the photodissociation 
cross section, exhibits a complicated behavior as function of the photon 
energy\cite{Dobbyn95}, behavior which suggests a statistical analysis of
the problem. This approach has been taken, recently, by Fyodorov and
Alhassid\cite{Fyodorov98,Alhassid98} who analyzed correlations of the  
photodissociation cross section in the framework of random 
matrix theory\cite{Mehta91}. In this framework, one is able to 
describe the universal properties of the  photodissociation 
cross section, but not the individual imprints of each molecule. 
It is well known, however, that the statistical properties of quantum 
chaotic systems bear a simple relation to the underlying classical 
dynamics\cite{Agam95}. The main purpose of this paper is to
relate the statistical  characteristics of the photodissociation 
cross section to correlation functions associated with the 
classical dynamics of the molecule.
 
Consider a molecule, in the ground state $|g\rangle$, excited
by a light pulse to an energy above  the dissociation threshold,
and let ${\cal H}$ denote the Hamiltonian of the system on the
excited electronic surface. It will be assumed that ${\cal H}$
represents an open system with several open dissociation channels. 
The photodissociation cross section of the molecule, in the 
dipole approximation, is given by
\begin{equation} 
\sigma(\epsilon)= - \eta~ \mbox{Im}~ \langle g | ~{\cal D} \frac{1}{ 
\epsilon^+ -{\cal H} } {\cal D} ~| g \rangle, \label{sigma} 
\end{equation} 
where $\epsilon$ is the energy 
measured from the ground state of the molecule and 
$\epsilon^\pm=\epsilon\pm i0$, ${\cal D} = {\bf d}\cdot \hat{\bf e}$ 
is the projection of the 
electronic dipole moment operator of the molecule, ${\bf d}$, on the 
polarization, $\hat{\bf e}$, of the absorbed light, and  $\eta=\epsilon/c 
\hbar \epsilon_0$,  $c$ being the speed of light, and $\epsilon_0$ the 
electric permitivity. 

We focus our attention on the dimensionless two-point correlation function: 
\begin{equation} 
K(\omega) =\frac{\left\langle  \sigma(\epsilon - \frac{\hbar \omega}{2} ) 
\sigma(\epsilon + \frac{\hbar\omega}{2} ) \right\rangle}{\langle   
\sigma(\epsilon)\rangle^2} -1, \label{def} 
\end{equation} 
where $\langle \cdots \rangle$ denotes an energy averaging over a classically
small energy interval centered at $\epsilon$ which, nonetheless, 
includes a large number of resonances. It will be assumed that 
the excitation energy,  $\epsilon$, is  sufficiently high such  
that the mean spacing between the vibrational modes of the molecule 
is smaller than the energy broadening due to the finite life time 
of the system in the excited states. This is the regime of 
overlapping resonances.  
 
Turning to the classical counterpart of our system, let $\rho({\bf x})$
be the initial density distribution of the excited molecule in phase space.
Here ${\bf x}= ({\bf r},{\bf p})$ is a phase space point, ${\bf r}$ and ${\bf 
p}$ being $d$ dimensional vectors of coordinates and conjugate momenta, 
respectively. The classical autocorrelation 
function of $\rho({\bf x})$ is defined as
\begin{equation} 
C(t) = \frac{\langle \langle \rho({\bf x}(t )) \rho({\bf x}) 
\rangle \rangle}{\langle \langle \rho({\bf x})\rangle \rangle^2}, 
\label{correlationf} 
\end{equation} 
where ${\bf x}(t)$ is the end point of the trajectory starting 
at ${\bf x}(0)={\bf x}$, and evolving for a time $t$ according to
the classical equations of motion: ${\bf \dot{r}}= \partial{\cal H}/\partial 
{\bf p}$ and ${\bf \dot{p}}= -\partial{\cal H}/\partial {\bf r}$. The 
averaging, $\langle \langle \cdots \rangle \rangle$, is over all initial 
points, ${\bf x}$, on the energy shell $\epsilon = {\cal H}({\bf x})$ 
(see definition in Eq. (\ref{micro}) below). Another classical correlation
function, closely related to $C(t)$, is defined by restricting the
average in (\ref{correlationf}) only to those points ${\bf x}$ which lie on 
the periodic orbits of the system with period $\tau$, where $\tau\geq t$. This 
function will be denoted by $C_\tau(t)$, and its precise definition 
will be given later on (see Eq.~(\ref{corr2})).

The central result of this paper is a formula which expresses
the two-point correlation function of the photodissociation cross section,
$K(\omega)$, in terms of the classical correlation functions, $C(t)$,
and $C_\tau(t)$. Namely, within the semiclassical approximation,
$K(\omega)$ takes the form
\begin{eqnarray}
K(\omega) \simeq K_1(\omega) + K_2 (\omega) , 
\label{sumf}
\end{eqnarray}
where
\begin{eqnarray}
K_1(\omega) = \frac{2}{\pi \beta \hbar} \mbox{Re} \! \int_0^\infty \!\! 
d t~e^{i\omega t} C(t), ~~~~~~\mbox{and}~  \label{k1} \\
K_2(\omega) = \frac{1}{\pi^2 \beta \hbar^2}\mbox{Re} \!\! 
\int_0^\infty\!\!\!\! 
d t~e^{i\omega t} \int_{-\frac{t}{2}}^{\frac{t}{2}}  \!\! d \tau
~C_t(\tau). 
\label{k2}
\end{eqnarray}
Here $\beta=1$ for Hamiltonians with time reversal symmetry, 
and $\beta =2$  for systems without this symmetry. It is assumed
that ${\cal H}$ have no other discrete symmetry. 

The above results will be derived by 
semiclassical methods\cite{Gutzwiller90}. 
The overlap of resonances implies that the behavior 
of $\sigma(\epsilon)$ is dominated by short time dynamics, and therefore the 
semiclassical approximation is justified. A semiclassical treatment of this 
problem will also clarify the role of various classical trajectories, and 
help interpreting formulae (\ref{k1}-\ref{k2}). 
 
In the semiclassical limit,  
the Green function is approximated by a sum of two  
terms: $G \simeq G_W + G_{osc}$. The first, known as the Weyl term, is 
a smooth function of the energy, $\epsilon$:  
\begin{equation} 
\langle {\bf r}' | G_W(\epsilon^\pm)  | {\bf r} \rangle =  
\int \frac{d {\bf p}}{(2 \pi \hbar)^d} 
\frac{e^{\frac{i}{\hbar} {\bf p\cdot} ({\bf r' - r})}} 
{\epsilon^\pm -{\cal H}\left( {\bf p}, 
\frac{{\bf r' + r}}{2} \right)}. \label{Weyl} 
\end{equation} 
It is large when $|{\bf r}-{\bf r'}| $ is of order of the particle wavelength, 
and therefore represents a local function in space. 
The second contribution, $G_{osc}$, is nonlocal in space 
but oscillatory in the energy. 
It is expressed as a sum over classical trajectories\cite{Berry72}: 
\begin{equation} 
\langle {\bf r}' | G_{osc}(\epsilon^\pm) |{\bf r} \rangle =  
\sum_{\nu} A_\nu^\pm e^{\pm \frac{i}{\hbar}  
S_\nu(\epsilon;{\bf r'},{\bf r})}, \label{osc} 
\end{equation}  
where $S_\nu(\epsilon;{\bf r'},{\bf r})$ is the classical action of the 
$\nu$-th trajectory from ${\bf r}$ to ${\bf r'}$ with energy $\epsilon$, 
and $A_\nu^+= (A_\nu^-)^*$ is the corresponding amplitude. 

This decomposition of the Green function implies that
the average, $\langle \sigma(\epsilon) \rangle$, comes only from the Weyl 
contribution, $G_W$. A straightforward calculation yields 
$\langle \sigma(\epsilon) \rangle =  \langle \langle 
\rho({\bf x}) \rangle \rangle $, where 
\begin{equation} 
\rho({\bf x}) = \int d{\bf r'} e^{\frac{i}{\hbar}{\bf p \cdot r'}} 
\langle {\bf r}- \frac{\bf r'}{2} | \psi \rangle \langle  \psi  
|{\bf r}+ \frac{\bf r'}{2}\rangle \label{wignerQ}
\end{equation}
is the Wigner function of the (unnormalized) state $|\psi \rangle = 
{\cal D} |g\rangle$. This function is real, and upon small smearing 
in phase space yields a positive definite function which can be 
interpreted as the initial density distribution of 
the excited molecule. The microcanonical average, 
$\langle \langle ~~~~\rangle \rangle$, of a general function $f({\bf x})$
is defined as 
\begin{equation} 
 \langle \langle f({\bf x}) \rangle\rangle =  \int 
\frac{d {\bf x}}{(2 \pi \hbar)^d} ~ 
\delta ( \epsilon - {\cal H} ({\bf x}) )f({\bf x}). \label{micro} 
\end{equation} 
Notice that $f$ and $\langle \langle f \rangle\rangle$ do not have
the same dimensions. To simplify our notations, from now on
we  work in units where $\hbar = 1$.

The correlator of $\sigma(\epsilon)$ can be written in the form: 
\begin{eqnarray} 
\left\langle \sigma(\epsilon \!-\! \frac{\omega}{2}) \sigma(\epsilon \!+\! 
\frac{\omega}{2} ) 
\right\rangle 
 =  \frac{\eta^2 }{2 \pi^2}\! \int \!\prod_{i=1}^4 d{\bf r}_i 
\rho({\bf r}_1,{\bf r}_2)  \rho({\bf r}_3,{\bf r}_4) \nonumber \\
\!\!\!\!\!\!\!\!\!\!\!\!\!\!\!\!\!\!\!\!\!\!\!\times \mbox{Re}
\left\langle G(\epsilon^-\!-\!\frac{\omega}{2}; {\bf r_2}, {\bf r_1})
G(\epsilon^+\!+\!\frac{\omega}{2}; {\bf r_4}, {\bf r_3})\right\rangle, 
\label{integral}
\end{eqnarray} 
where $\rho({\bf r},{\bf r'}) = \langle {\bf r} | \psi \rangle \langle \psi | 
{\bf r'}\rangle$. 
Considering the connected part of the correlator, only the oscillatory  
terms of the Green functions, $G_{osc}$, contribute.  Since these are strongly 
oscillating functions of the coordinates, the main contribution to the 
integral (\ref{integral}) comes from two types of orbits: (a) general orbits 
with initial and final points far apart, but  ${\bf r_2} \sim {\bf r_3}$ and  
${\bf r_1} \sim {\bf r_4}$. (b) Returning orbits with ${\bf r_1} \sim 
{\bf r_2}$ and  ${\bf r_3} \sim {\bf r_4}$. The corresponding contributions to 
(\ref{integral}) will be denoted by $I_a$ and $I_b$, respectively.  
When ${\cal H}$ is also time 
reversal symmetric,  additional contribution, equal to $I_a$, comes from 
orbits with ${\bf r_1} \sim {\bf r_3}$ and  ${\bf r_2} \sim {\bf r_4}$. 
  
The calculation of $I_a$ involves two main approximations. One is to expand 
the actions of the Green functions (\ref{osc})
to the first order in $\omega$, ${\bf r}_3 - {\bf r}_2$, and 
${\bf r}_4 - {\bf r}_1$. For this purpose we use the relation: 
$S_\nu(\epsilon \pm \omega/2 ; {\bf r}'+\delta {\bf r}', {\bf r} + 
\delta {\bf r}) \simeq S_\nu(\epsilon; 
{\bf r}', {\bf r}) \pm t_\nu \omega/2 + {\bf p'_\nu} \cdot \delta {\bf r}' - 
{\bf p_\nu} \cdot  \delta {\bf r}$, where $t_\nu$  
is the time of the orbit, ${\bf p_\nu}$ is the initial momentum of the 
trajectory (at point ${\bf r}$), and ${\bf p'_\nu}$ is the final 
momentum (at point ${\bf r}'$). The second approximation is to neglect 
quantum interference effects (weak localization corrections) and employ 
the ``diagonal approximation". In this approximation, the double sum on 
trajectories in the product of Green functions, $G_{osc}(\epsilon^+) 
G_{osc}(\epsilon^-) \simeq \sum_{\nu \nu'} A_\nu A_{\nu'}^* e^{i S_\nu - 
i S_{\nu'}}$, is replaced by a single sum in which one keeps only pairs 
of trajectories, $(\nu ,\nu')$, related by symmetry.
Noticing that $\rho({\bf r}_1,{\bf r}_2) \rho({\bf r}_3,{\bf r}_4) = 
\rho({\bf r}_1,{\bf r}_4) \rho({\bf r}_3,{\bf r}_2)$, and integrating on the 
coordinate differences ${\bf r}_3 - {\bf r}_2$, and ${\bf r}_4 - {\bf r}_1$, 
one thus obtains: 
\begin{equation} 
I_a \simeq \frac{\eta^2}{2 \pi^2} \mbox{Re} \int   d{\bf r} d{\bf r'}  
\sum_\nu |A_\nu|^2\rho({\bf x'_\nu}) \rho({\bf x_\nu})e^{it_\nu \omega }, 
\label{I1} 
\end{equation} 
where ${\bf x_\nu}=({\bf r},{\bf p_\nu})$ and ${\bf x'_\nu}= 
({\bf r}',{\bf p'_\nu})$ are the initial and final phase-space coordinates 
of the $\nu$-th trajectory.

Now, to evaluate the sum over $\nu$, we use the sum rule\cite{Agam96}: 
\begin{eqnarray} 
(2 \pi)^{d-1} \sum_\nu |A_\nu|^2 g({\bf x'_\nu}, {\bf x_\nu}, t_\nu) =
~~~~~~~~~~~~~~~~~~~~~~~ \label{sumrule} \\
~~~ 
\int_0^\infty \! dt d{\bf p}' d{\bf p} ~ g({\bf x}', {\bf x}, t)  
\delta(\epsilon -{\cal H} ({\bf x})) \delta ({\bf x}'-{\bf x}(t)),
\nonumber 
\end{eqnarray} 
where $g({\bf x}', {\bf x}, t)$ is a general function of the  
time $t$, the initial ${\bf x}$, and the final ${\bf x}'$ phase space points. 
${\bf x}(t)$ is the phase space coordinate, after time $t$,  of a trajectory 
starting from ${\bf x}(0)={\bf x} $.   
Eq. (\ref{sumrule}) can be proved by performing the integrals on its 
right hand side. For this purpose it is convenient to use a local 
coordinate system with one coordinate parallel to the orbit, and
$d-1$ perpendicular to it\cite{Agam96}. 
Substituting (\ref{sumrule}) in (\ref{I1}), and 
integrating over ${\bf x}$  
yields 
\begin{equation} 
I_a \simeq \frac{2 \eta^2}{(2 \pi)^{d+1} } \mbox{Re} \!\!
 \int_0^\infty \!\!\!\!d t \!\int \!\!d{\bf x}~  
e^{it\omega }  
\delta (\epsilon \!-\! {\cal H}({\bf x})) \rho({\bf x}(t)) \rho({\bf x}). 
\end{equation}  
Using the definition of microcanonical average (\ref{micro}) and the  
correlation function (\ref{correlationf}), one obtains 
$K_1(\omega) = 2 I_a/\langle \langle \rho \rangle \rangle^2 \beta$, 
where $\beta$ is the symmetry factor 
associated with the additional contribution in the case of time reversal 
symmetry.  

To evaluate $I_b$, the term associated with returning trajectories, 
we expand the actions of the Green functions in (\ref{integral}), 
to first order in  
${\bf r}_2 -{\bf r}_1$  and ${\bf r}_4 -{\bf r}_3$.  
Integrating over these coordinate differences, the result takes the form  
$I_b= \eta^2 \mbox{Re} \left\{ I(\omega) I^*(\omega)\right\}/2 \pi^2$, 
where  
\begin{equation} 
I(\omega) \simeq \int d{\bf r} \sum_\nu A_\nu \rho({\bf x_\nu}) 
e^{i S_\nu(\epsilon +
\frac{\omega}{2};{\bf r},{\bf r})}. \label{I+}
\end{equation} 
Here ${\bf x_\nu}=({\bf r},\overline{{\bf p}}_\nu)$ is a phase space point
corresponding to the $\nu$-th returning 
trajectory at point ${\bf r}$, where $\overline{{\bf p}}_\nu =
({\bf p_\nu'}+{\bf p_\nu})/2$ is the average of the 
initial, ${\bf p_\nu}$,  and final, ${\bf p'_\nu}$, momentum of the 
trajectory.
The integral over ${\bf r}$ is now performed in the stationary 
phase approximation. The stationary phase condition,
$\partial  S_\nu(\epsilon;{\bf r},{\bf r})/\partial 
{\bf r}={\bf p_\nu}-{\bf p_\nu'}=0$, implies that the initial and 
final momentum of the orbit are equal, therefore, the main contribution 
to the integral (\ref{I+}) comes from the vicinities of periodic 
orbits\cite{Gutzwiller67}. The result  
can be expressed as a sum over periodic orbits: $I(\omega)= 
\sum_p A_p \rho_p e^{iS_p(\epsilon)+ 
i t_p \omega/2}$, where $A_p$, $t_p$ and $S_p$ are the
amplitude, period, and action of the $p$-th periodic orbit respectively. 
$\rho_p = \oint dt \rho({\bf x}_p(t))$ is the time integral of 
$\rho({\bf x})$ along the orbit ${\bf x}_p(t)$. 

Next, we analyze the product $I(\omega) 
I^*(\omega)$, which in the diagonal approximation is given by
$(2/\beta) \sum |A_p|^2 \rho_p^2 e^{i t_p \omega}$. 
Following Eckhardt et al.\cite{Eckhardt95},
\begin{eqnarray}
\rho_p^2 = \int_0^{t_p} d\tau\int_0^{t_p} d\tau' 
\rho({\bf x}_p(\tau))\rho({\bf x}_p(\tau')) =~~~~~~ 
\nonumber\\
\int_0^{t_p} dt \int_{-\frac{t_p}{2}}^{\frac{t_p}{2}} dt' 
\rho({\bf x}_p(t-\frac{t'}{2}))
\rho({\bf x}_p(t+\frac{t'}{2})) = ~~~~~\nonumber \\
= t_p \int_{-\frac{t_p}{2}}^{\frac{t_p}{2}} dt' \left\langle  
\rho({\bf x}(t')) \rho({\bf x}) \right\rangle_p, ~~~~~~~~~~~~\nonumber
\end{eqnarray}
where  $\langle (\cdots) \rangle_p= t_p^{-1}\oint dt (\cdots )$ denotes the 
time average along the $p$-th periodic orbit. Thus $\langle I(\omega) 
I^*(\omega)\rangle =(2/\beta) \sum t_p |A_p|^2  e^{i t_p \omega}
\langle  \rho({\bf x}(t')) \rho({\bf x}) \rangle_p $. The sum over the
periodic orbits can be converted into an integral over time using 
the relation
\begin{eqnarray}
\sum_{t<t_p<t+\delta t} \!\!\!\!\!t_p |A_p|^2 \langle  ~\cdots ~ 
\rangle_p = ~~~~~~~~~~~~~~~~~~~~~~~~~~~~~~~ \label{posum} \\
~~~=\delta t \int \!\!d{\bf x} \delta [\epsilon \!-\!{\cal H}({\bf x})]
\delta [{\bf x}_\parallel\! - \!{\bf x}_\parallel(t)] ( \cdots ),
 \nonumber
\end{eqnarray}
where ${\bf x}_\parallel$ is a phase space coordinate on the energy shell
${\cal H}({\bf x})=\epsilon$, while ${\bf x}_\parallel(t)$ is the 
position of the system after time $t$, starting from
${\bf x}_\parallel(0)={\bf x}_\parallel$.
Using (\ref{posum}), 
the sum over periodic orbits is converted into a time 
integral which yields formula (\ref{k2}) for $K_2(\omega)$, 
with the correlation function:
\begin{eqnarray}
C_\tau(t)=
 \int \!\! d {\bf x}\frac{\rho({\bf x}(t)) \rho({\bf x})}{ \langle 
\langle \rho \rangle \rangle^2 } \delta [\epsilon \!-\!{\cal H}({\bf x})]
\delta [{\bf x}_\parallel\! - \!{\bf x}_\parallel(\tau)]. \label{corr2}
\end{eqnarray}

The identification of the classical trajectories which underly the main 
contribution to the correlator $K(\omega)$, clarifies the origin
of fluctuations in $\sigma(\epsilon)$. These have two sources: 
the wave functions of the system, and the 
density of states. $K_1(\omega)$, associated with general trajectories, 
is related to fluctuations in the wave functions, while $K_2(\omega)$, 
coming from the periodic orbits of the system, is 
related to fluctuations in the density of states. 

To proceed further, one need to characterize the behavior of the correlation
functions $C(t)$ and $C_\tau(t)$. For this purpose we assume that the classical
dynamics of the system is characterized by two time scales which are well 
separated. Such a situation appears, for example, in almost close 
chaotic systems. There the long time scale is the typical 
time for dissociation, 
while the short time scale is the time which takes for a classical density 
distribution to relax to the ergodic distribution on the energy shell. 
We denote by ${\cal H}_0$ the closed Hamiltonian which controls the dynamics
of the system for time shorter than the dissociation time. 
Under these assumptions
$C(t) \simeq \Delta C_t(\tau) \simeq e^{-\gamma |t|}/\Delta$,
where $\gamma$ is the dissociation rate, and $\Delta$ is the mean spacing 
between resonances, i.e. $(2 \pi \hbar)^d/\Delta = 
\int d {\bf x} \delta ( \epsilon - {\cal H}_0 
({\bf x}) )$. We assume $\Delta$ to be approximately constant within the 
interval of energy averaging, and from now on work in units 
where $\Delta = 1$. If $\Delta$ is not constant, 
$\sigma(\epsilon)$ should be unfolded appropriately.

An infinite separation between the time scales mentioned above 
corresponds to the limit of random matrix theory\cite{Mehta91}. 
In this case $K(\omega)$ reduces to
\begin{equation}
K_u(\gamma,\omega) \simeq \frac{2}{\beta \pi} \left( \frac{\gamma}{\omega^2 + 
\gamma^2} + 
\frac{1}{2 \pi} \frac{ \gamma^2 -\omega^2}{(\gamma^2 + \omega^2)^2} \right). 
\label{universal}
\end{equation}
The first and the second terms of this formula come from  $K_1(\omega)$ 
and $K_2(\omega)$ respectively. As can be seen,
these components correspond to the leading and the subleading terms 
of an expansion in $1/\gamma$.

Formula (\ref{universal}), describing the correlations 
in the universal regime, was first derived by Fyodorov and 
Alhassid\cite{Fyodorov98} using the nonlinear $\sigma$-model\cite{Efetov83}.
Here we confirm their conjecture that for overlapping resonances ($\gamma >1$)
$K(\omega)$ can be derived by semiclassical methods. 

Yet, the range of validity of formulae (\ref{sumf}-\ref{k2}) 
goes far beyond the universal regime. They account also for system 
specific contributions to $K(\omega)$, which might  be of the 
same order of the universal result (\ref{universal}).
To be concrete, we focus on the nonuniversal corrections due to the leading
term, $K_1(\omega)$, and consider a representative behavior of decay of
correlations in open chaotic systems:
\begin{equation}
C(t) \simeq e^{\gamma | t|} +  
\alpha e^{- \gamma_2 |t |} \cos (\omega_2 t). \label{decay}
\end{equation}
Here $\gamma_2$ is the real part of the second Ruelle 
resonance\cite{Ruelle86} of the classical system, while $\omega_2$ is 
its imaginary part. $\alpha$ is a constant of order unity which depends 
on details of the system and the initial state (\ref{wignerQ}). 
The situation where $\omega_2 \neq 0$ appears usually
when a specific short periodic orbit has strong influence on the 
dynamics of the system, since every typical trajectory 
stays in its vicinity for a long time.
Thus, $\omega_2 \approx 2 \pi/t^*$,  where $t^*$ is the period 
of the orbit. Substitution of (\ref{decay}) in (\ref{sumf}-\ref{k2}) yields:
\begin{equation}
K(\omega)\simeq K_u(\gamma,\omega) + \frac{\alpha}{\pi \beta} \sum_\pm
\frac{\gamma_2}{\gamma_2^2+ (\omega \pm \omega_2)^2}.
\label{non}
\end{equation}
{\narrowtext
\begin{figure}[h]
\epsfxsize=9cm
\vspace{-0.5 cm}
\epsfbox{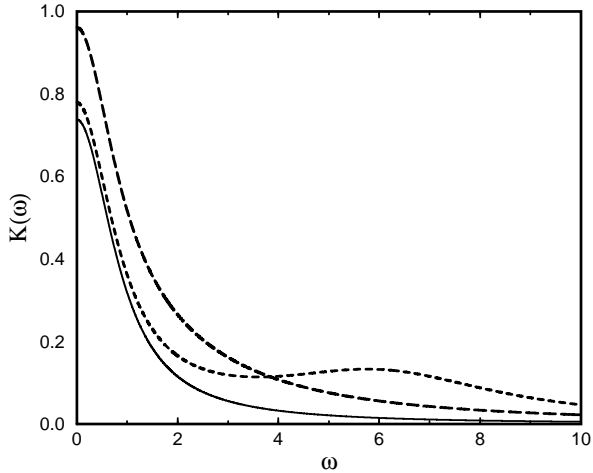}
\caption{The two-point correlation function of the photodissociation 
cross section for systems with time reversal symmetry ($\beta = 1$).
The solid line is the universal result $K_u(\gamma, \omega)$  (Eq.~18) 
for the case $\gamma = 1$. The dashed and the dotted lines correspond to the  
correlation function, $K(\omega)$, which includes the nonuniversal 
contribution, in the approximation of Eq. (20). In both cases 
$\gamma=1$, $\gamma_2=3$ and $\alpha=1$, but, the dashed line 
corresponds to $\omega_2=0$, while the dotted line to $\omega_2=6$.}
\label{fig:1}
\end{figure}
}
\noindent
This formula is plotted for various parameters in Fig.~1. The solid line 
represents the universal form (\ref{universal}). The dashed line,
corresponding to $\omega_2= 0$, shows the typical behavior of systems 
where the decay of correlations is of diffusive nature. 
In this case the main deviation from universality 
appears as an increase of correlations near the peak of $K(\omega)$, 
at $\omega =0$. The dotted line corresponds to a nonzero value of 
$\omega_2$, and characterizes the behavior of ballistic systems. 
Here the nonuniversal contribution is located in the tail of $K(\omega)$,
near $\omega=\omega_2$, where the universal term (\ref{universal}) is 
already negligible. These plots demonstrate the significance of 
system specific contributions to $K(\omega)$. 

In conclusion, we derived the semiclassical relation between 
the two point correlation function, $K(\omega)$,
and the classical correlation functions $C(t)$, and $C_\tau(t)$. 
The first quantity characterizes the quantum mechanical process 
of photodissociation, while $C(t)$, and $C_\tau(t)$ 
characterizes the dynamics of the
classical counterpart of the system. In contrast with the two-point
correlation function of the density of states, where periodic orbits play 
the major role, here we have found that the main contribution to 
$K(\omega)$ comes from general orbits. Moreover, the nonuniversal 
contributions to the correlator, $K(\omega)$, can be of the same order of
the universal result of random matrix theory (the relative
magnitude being proportional to $\gamma/\gamma_2$). These results can be 
used in order to analyze the experimental data of complex molecules, 
extract information regarding their classical dynamics, and construct 
effective models for these molecules. 

I would like to thank Yan Fyodorov for pointing out the problem of the 
semiclassical derivation of formula (\ref{universal}), and to Shmuel Fishman, 
Raphy Levine, and Nadav Shnerb for useful discussions and comments.
This work was initiated at the ``Extended Workshop on Disorder, 
Chaos and Interaction in Mesoscopic System'' which took place in Trieste 
1998. I thank the I.C.T.P. for the generous hospitality.

\end{multicols}  
\end{document}